\documentclass[aps,prl,twocolumn,superscriptaddress,oneside,floatfix,amsmath,showpacs,amssymb]{revtex4-1}
\usepackage{graphicx}
\usepackage{dcolumn}
\usepackage{bm}

\begin{document}

\title{Decrypting the cyclotron effect in graphite using Kerr rotation spectroscopy}

%
%

\author{Julien Levallois}
\affiliation{D\'{e}partement de Physique de la Mati\`{e}re Condens\'{e}e,
Universit\'{e} de Gen\`{e}ve, CH-1211 Gen\`{e}ve 4, Switzerland}

\author{Micha\"{e}l Tran}
\affiliation{D\'{e}partement de Physique de la Mati\`{e}re Condens\'{e}e,
Universit\'{e} de Gen\`{e}ve, CH-1211 Gen\`{e}ve 4, Switzerland}

\author{Alexey B. Kuzmenko}
\affiliation{D\'{e}partement de Physique de la Mati\`{e}re Condens\'{e}e,
Universit\'{e} de Gen\`{e}ve, CH-1211 Gen\`{e}ve 4, Switzerland}

\date{\today}

\begin{abstract}
We measure the far-infrared magneto-optical Kerr rotation and reflectivity
spectra in graphite and achieve a highly accurate unified microscopic
description of all data in a broad range of magnetic fields by taking rigorously the
c-axis band dispersion and the trigonal warping into account. We find that
the second- and the forth-order cyclotron harmonics are optically almost as
strong as the fundamental resonance even at high fields. They must play,
therefore, a major role in magneto-optical and magneto-plasmonic applications
based on Bernal stacked graphite and multilayer graphene.
\end{abstract}

 \maketitle

Owing to its wide spread and technological importance, graphite is one of the
most studied crystalline materials. Magneto-optical spectroscopy was one of
the key techniques, together with the transport and magnetization
measurements, to establish the essentials of the unusual band structure of
this layered semimetal. A small mass, low density and weak scattering of
electrons and holes give rise to remarkably strong cyclotron
resonances~\cite{GaltPR56,SuematsuJPSJ72,ToyPRB77,DoezemaPRB79,NakamuraJPSJ84,HofmannRSI06,LiPRB06,OrlitaPRL08,ChuangPRB09,UbrigPRB11,TungCM11}.
Surprisingly, in spite of the large number of experiments that appeared after
the first report of the cyclotron effect~\cite{GaltPR56}, the
magneto-optical lineshapes and intensities are still not satisfactorily
explained by microscopic calculations, which is in part because of the
incompleteness of the existing optical data. While the measurements where
circular polarization served to distinguish electrons and holes were
typically done with only a few radiation
wavelengths~\cite{SuematsuJPSJ72,ToyPRB77,DoezemaPRB79,NakamuraJPSJ84}, most
broadband spectroscopic
experiments~\cite{LiPRB06,OrlitaPRL08,ChuangPRB09,TungCM11} provide a highly
mixed response of electrons and holes. As there exist several open questions
regarding the optical intensities in
graphene~\cite{GusyninPRL07,LiNP08,CrasseePRB11}, a quantitative
understanding of graphite as its three-dimensional counterpart is highly
desirable.

In order to combine the advantages of broadband spectroscopy and a
sensitivity to the charge carrier type, we measure the Kerr rotation angle
$\theta_{K}(\omega)$ as a continuous function of the photon energy
$\hbar\omega$, in addition to the more conventional reflectivity spectra
taken on the same sample. The Kerr angle is defined as the rotation of the
polarization plane upon a normal reflection of linearly polarized light from
the surface of a sample, when the magnetic field $B$ is perpendicular to the
surface. The reflectivity is a ratio of the intensities of the reflected and
the incident beams of unpolarized radiation in the same geometry. These
quantities are related to the complex reflectivity coefficients
$r_{\pm}(\omega)$ for the right (+) and left (-) circular polarized light:
\begin{equation}
\theta_{K}=\frac{\mbox{arg}(r_{-}) - \mbox{arg}(r_{+})}{2},
R=\frac{|r_{-}|^2+|r_{+}|^2}{2} \label{EqThetaR}
\end{equation}
\noindent and eventually determined, via the usual Fresnel
equations, by the circular conductivities
$\sigma_{\pm}(\omega)=\sigma_{xx}(\omega)\pm i
\sigma_{xy}(\omega)$. While the electron and hole cyclotron
resonances are indistinguishable in the reflectivity, they can
be separated by the Kerr rotation. The latter has the same
physical origin as the Faraday rotation~\cite{CrasseeNP11}, but
can be applied to opaque samples. This set of measurements can
be regarded as infrared Hall spectroscopy~\cite{KaplanPRL96}
since it probes both the longitudinal and the Hall
conductivities, $\sigma_{xx}$ and $\sigma_{xy}$, at infrared
frequencies as specified in the Supplemental Material.

Figures 1a and 1b show the Kerr angle and the reflectivity of highly ordered
pyrolytic graphite (HOPG) at 5~K for magnetic fields up to 7~T. The Kerr
spectra are complicated and highly structured above 1~T; the structures
displace in energy roughly proportionally to the field. They crossover to a
simpler shape at lower fields before disappearing, as expected, at
$B\rightarrow 0$. At higher fields the quantized Landau level (LL)
transitions can be resolved, while at low fields different transitions merge
and form a broad continuum. The reflectivity demonstrates a similar trend,
except that it remains finite at zero field, and this agrees with previously
taken magneto-reflectivity spectra at the corresponding photon energies and
magnetic fields~\cite{LiPRB06,TungCM11}.

\begin{figure}
\centering
\includegraphics[width=9cm]{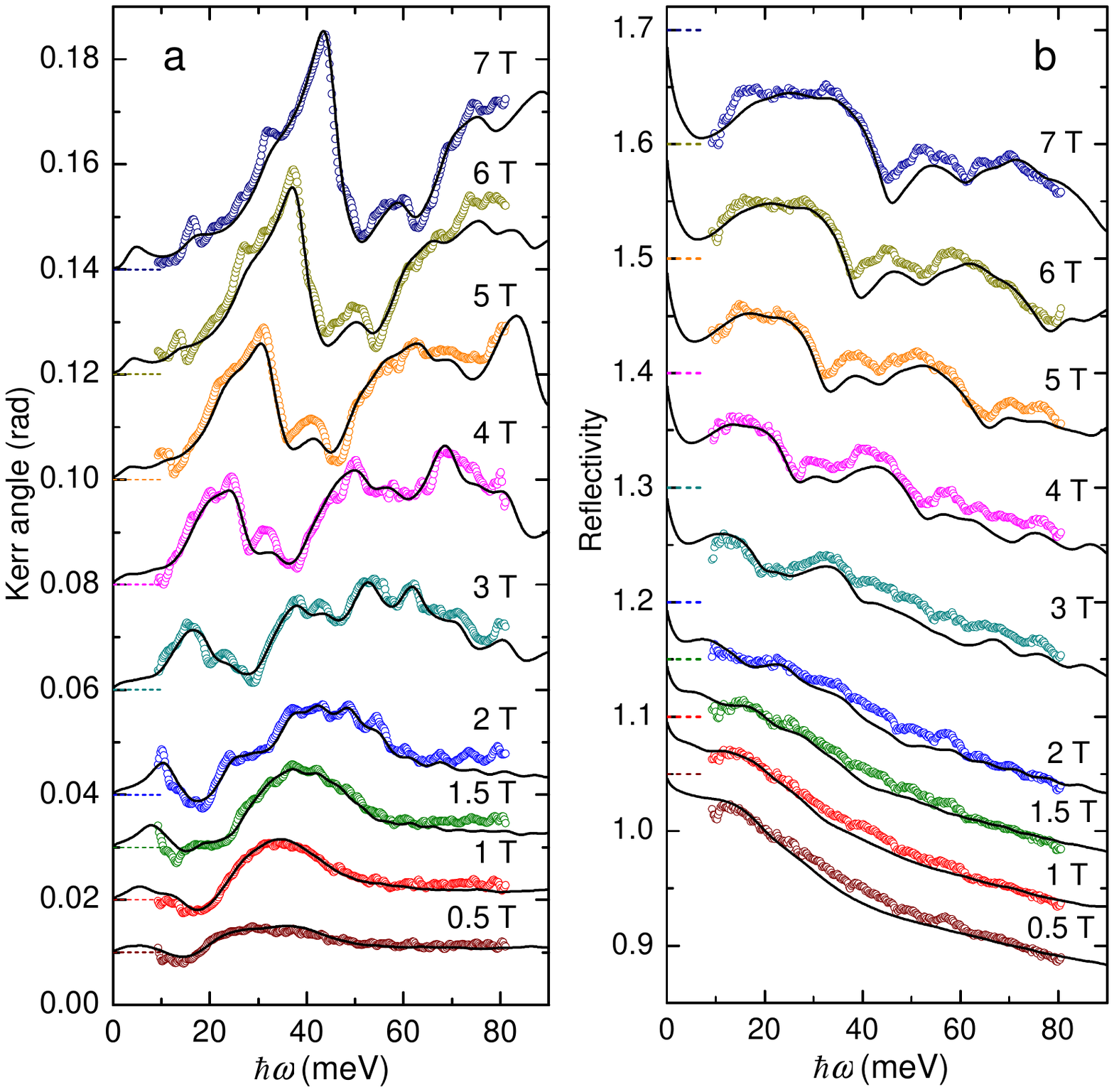}\\
\caption{(color online) The Kerr angle (a) and reflectivity (b) spectra of
highly ordered pyrolytic graphite measured at 5 K at various
magnetic fields. The curves are offset for clarity; the dashed
lines indicate 0 at panel (a) and 1 at panel (b).
Circles are the experimental data, solid lines are the
theoretical fits using the same set of tight-binding parameters
at all fields. Note that only the Kerr angle was used to adjust the band parameters.} \label{Fig1}
\end{figure}

\begin{figure*}
\centering
\includegraphics[width=14cm]{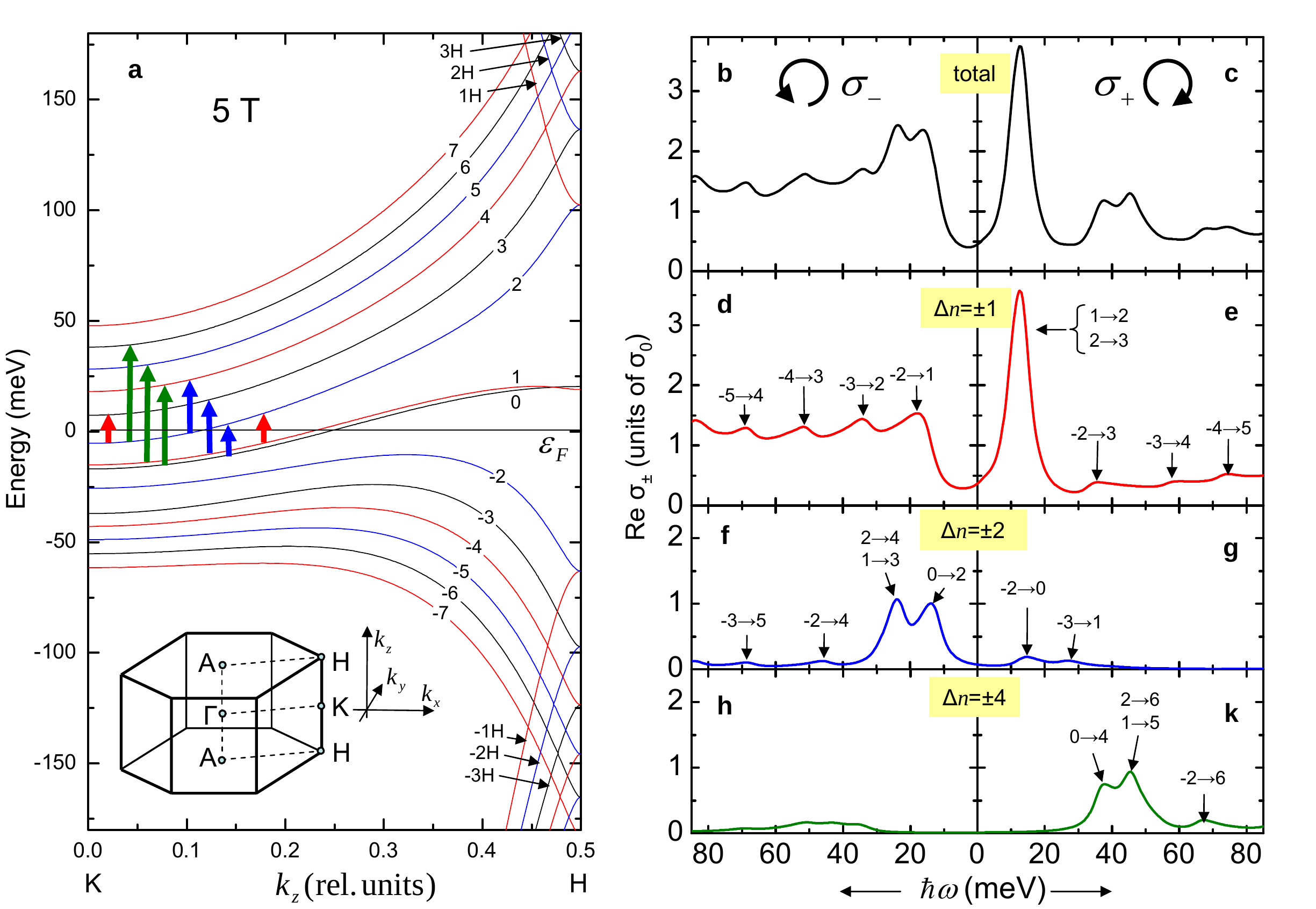}\\
\caption{(color online) Panel (a): calculated low-energy Landau-level structure
of graphite at 5~T. The inset shows the Brillouin zone of graphite. (b)-(k): The
theoretical spectra of Re $\sigma_{+}(\omega)$ (right panels) and Re
$\sigma_{-}(\omega)$ (left panels) at 5~T and 5~K. b, c: the total
conductivity; (d), (e): the contribution from the fundamental transitions $\Delta
n = \pm 1$; (f), (g): the contribution from the high-order transitions $\Delta n
= \pm 2$; (h), (k): the same for $\Delta n = \pm 4$. The horizontal axis for the
left panels is inverted for the illustration purpose, since Re
$\sigma_{+}(-\omega)$ = Re $\sigma_{-}(\omega)$. Conductivities per atomic
layer are shown in the units of $\sigma_0=e^2/4\hbar$. The strongest
intraband transitions corresponding to the selection rules $n \rightarrow
n+1$, $n \rightarrow n+2$ and $n \rightarrow n+4$ are shown in panel (a) by the
red, blue and green arrows correspondingly.} \label{Fig2}
\end{figure*}

The extremely small Fermi surface of graphite is elongated along the K-H line
at the edge of the Brillouin zone (shown in the inset of Figure 2a),
perpendicular to the graphene planes. Between K and H, the in-plane bands
change continuously, the details being strongly dependent on the relative
strength of the different interplane hopping parameters~\cite{McClurePR57,SlonczewskiPR58}. However, it is often assumed in the literature that the observable spectral features originate only from the
K-point, where the Fermi surface is electron-like, and from the H-point,
where it is hole-like. An appealing aspect of this simplified view is that
the band structure for the H-point strongly resembles the conical band
dispersion of monolayer graphene, while the bands at the K point disperse
parabolically similar to bilayer graphene. The LLs and therefore the
cyclotron frequencies at these points are contrasted by their dependence on
the perpendicular magnetic field: $\sim\sqrt{B}$ and $\sim B$
respectively~\cite{ToyPRB77,OrlitaPRL08}.

We found that it is impossible to reproduce the measured optical curves using
this two-point approximation, \emph{i.e. }by considering a weighted
superposition of the theoretical optical spectra of monolayer and bilayer
graphene~\cite{AbergelPRB07}. In contrast, taking the entire Fermi surface
into account resulted in excellent fits, shown in Figures 1a and 1b as solid
lines. Within this approach, the total optical conductivity is given by an
integral over the K-H line~\cite{NakamuraJPSJ84, FalkovskyPRB11}
\begin{equation}
\sigma_{\pm}(\omega)=\int_{K}^{H}
\tilde{\sigma}_{\pm}(k_z,\omega) d k_{z}
\end{equation}
\noindent where $\tilde{\sigma}_{\pm}(k_z,\omega)$ is the partial
contribution from all optically allowed transitions between LLs at a given
perpendicular momentum $k_z$. In order to compute the LLs and optical matrix
elements, we used only the conventional tight-binding parameters
$\gamma_0,...,\gamma_5$ and $\Delta$ of the well known
Slonczewski-Weiss-McClure (SWMcC) band
model~\cite{McClurePR57,SlonczewskiPR58} and the scattering rate $\Gamma$,
which we assumed, for the sake of simplicity, to be energy-, momentum- and
magnetic field-independent. We took the SWMcC values from the recent
magnetotransport measurements~\cite{SchneiderPRL09} and fine-tuned them to
fit our Kerr spectra at four field values 1, 3, 5 and 7~T
\emph{simultaneously}, thus covering small fields and high fields (almost up
to the quantum limit~\cite{LiPRB06,ZhuNP09}) at the same time. Remarkably,
the Kerr spectra at other fields as well as all reflectivity spectra were
reproduced satisfactorily without extra fitting. The resulting parameters are
fairly consistent with previous studies~\cite{SchneiderPRL09,GruneisPRB08}.
Thus in the present case the Kerr spectra \emph{alone} are, in principle,
sufficient to determine the electronic band parameters. The details of the
calculations and the fits are given in the Supplemental Material.

One can notice that some of the LL structures, especially at low frequencies,
are sharper in the experiment than in the fitting curves. This is likely due
to our assumption that the scattering rate ($\Gamma \approx 2.8$~meV was
found by the least-squares technique) is constant. Indeed, reducing $\Gamma$
makes the overall fit worse, but it improves the match for some spectral
features.

Like it is in bilayer graphene, each LL in graphite (except for the first
two) is subdivided into four sublevels, corresponding to one of the four
$\pi$-bands, two electron- and two hole-like. However, in contrast to
graphene, the levels in the three-dimensional graphite acquire a $k_z$
dispersion~\cite{McClurePR57,NakaoJPSJ76,FalkovskyPRB11}, shown in Figures 2a
for 5~T. We adopt the
notation~\cite{AbergelPRB07,OrlitaPRL08,FalkovskyPRB11}, where the main LL
index $n$ starts from zero, and the minus sign is used to denote the hole
sublevels. Additionally, we use the symbol H after the index to distinguish
the high-energy sublevels. The LLs with $n=0$ and $n=1$ are special as they
have only one and three sublevels respectively.

Although the bands and the LLs depend on all the SWMcC parameters, the
symmetry of the wavefunctions and the optical selection rules are determined
by the interplane next-nearest neighbor hopping $\gamma_3\approx 0.3$~eV. It
is responsible for the so-called trigonal
warping~\cite{NozieresPR58,InoueJPSJ62}, which breaks the in-plane radial
symmetry of the bands. The case of zero warping is much easier to treat
analytically, since the entire Hamiltonian in a magnetic field is factorized
into a series of uncoupled $4\times4$ blocks for each LL. If $\gamma_3\neq 0$
then one has to diagonalize numerically an infinite
matrix~\cite{NakaoJPSJ76}. This case, however, is most interesting because it
comprises nontrivial effects such as the magnetic breakdown in
graphite~\cite{NakaoJPSJ76} and Lifshitz transitions in bilayer graphene as a
function of strain~\cite{MuchaPRB11}. Furthermore, trigonal warping was shown
to have a profound effect on the charge transport in weakly doped bilayer
graphene~\cite{NovoselovScience11}.

Without trigonal warping, only the transitions where the LL index $n$ changes
by $\pm$1 are allowed, due to the harmonic-oscillator-like structure of the
LL wavefunctions. Warping mixes LLs separated by 3 times an integer number
(\emph{i.e. }levels shown by the same color in Figure 2a) and makes also the
high-order resonances $\Delta n = 3m\pm1$ ($m = \pm1, \pm2...$) optically
active~\cite{NozieresPR58}. These cyclotron harmonics were indeed
observed~\cite{GaltPR56,SuematsuJPSJ72,DoezemaPRB79,NakamuraJPSJ84,LiPRB06}
and served~\cite{InoueJPSJ62,SuematsuJPSJ72,DoezemaPRB79} to determine
$\gamma_3$. However, little is known about the absolute optical strength of
these transitions. The very good match between the data and the theory allows
us to clarify this issue.

Figures 2b and 2c show the real parts of the total right and left-handed
optical conductivities as given by the model at 5~T and 5~K. The curves are
normalized to $\sigma_0 = e^2/4\hbar$, since the conductivity of graphite in
zero field is close to this universal value in a broad energy
range~\cite{KuzmenkoPRL08}. The net contributions from the transitions
satisfying the fundamental selection rule $\Delta n = \pm 1$ are presented
separately in Figures 2d and 2e, where also an assignment is given to every
peak. The most intense resonance at about 13~meV in $\sigma_{+}(\omega)$ is
due to the electron-like LL intraband transitions of the type $n\rightarrow
n+1$. The actual number of the transitions involved depends on the magnetic
field; in particular two of them ($1\rightarrow 2$ and $2\rightarrow 3$) are
activated at 5~T. They stem from the range of $k_z$ starting from K until
approximately the midpoint between K and H. This peak is somewhat analogous
to the cyclotron resonance in usual semiconductors. Additionally, a series of
interband peaks of the type $-n\rightarrow n+1$ in $\sigma_{+}(\omega)$ and
$-n-1\rightarrow n$ in $\sigma_{-}(\omega)$ are
present~\cite{KoshinoPRB08,FalkovskyPRB11,OrlitaPRL08}. The values of $k_z$
involved in these optical transitions span the entire Brillouin zone.

\begin{figure}[t]
\centering
\includegraphics[width=8cm]{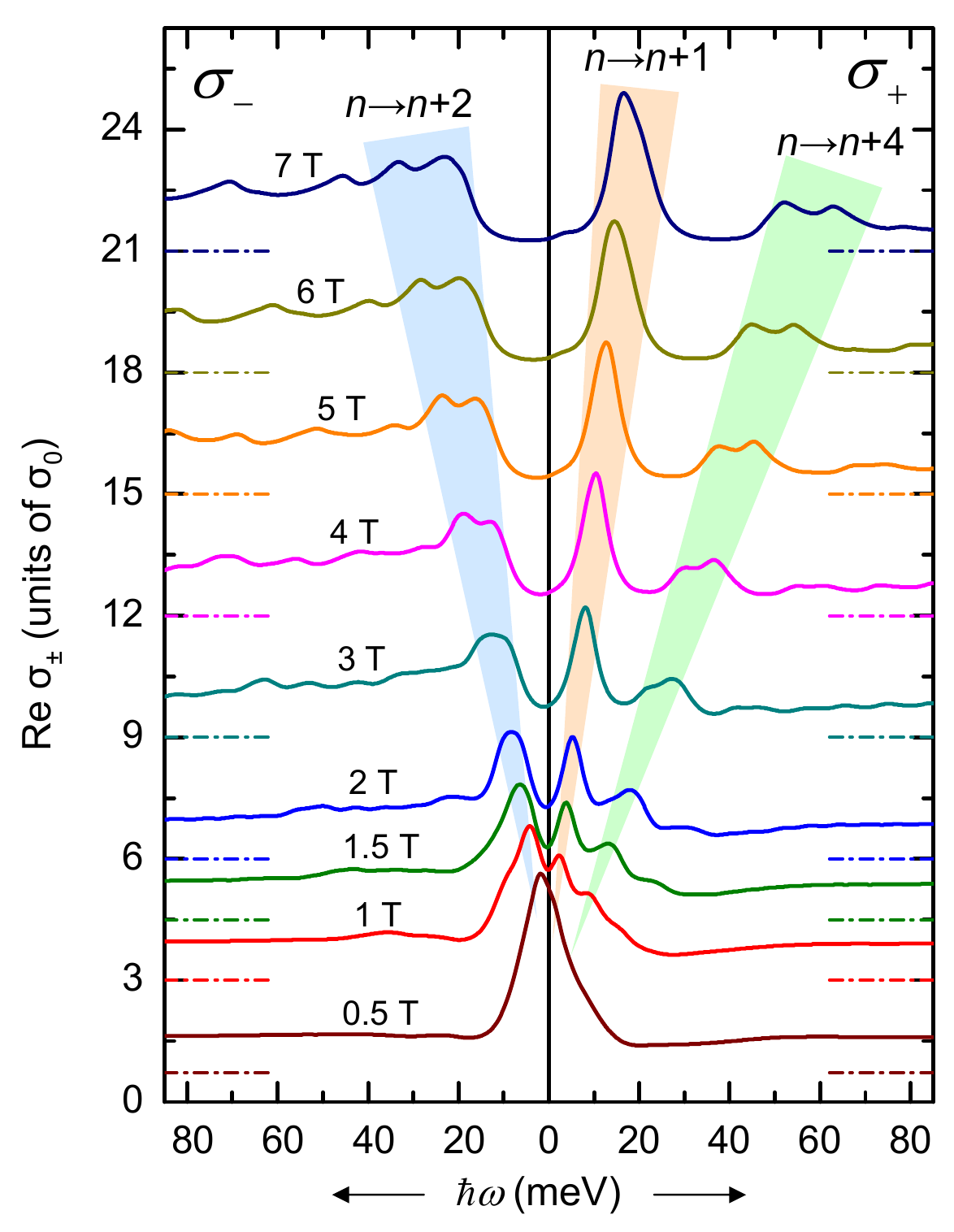}\\
\caption{(color online) The theoretical spectra of Re $\sigma_{-}(\omega)$ (on
the left) and Re $\sigma_{+}(\omega)$ (on the right) at 5~K at
the experimental values of magnetic field. In the calculations,
the same parameters were used that give the best fit to the
data. The curves are shifted for clarity; the zero levels are
shown by dashed-dotted lines of the same color. The peaks
corresponding to the fundamental cyclotron resonance $n
\rightarrow n + 1$ and the most intense harmonics $n
\rightarrow n + 2$ and $n \rightarrow n + 4$ are marked by
shaded regions.} \label{FigSigmaPM}
\end{figure}

The calculated partial optical conductivities due to the high-order
resonances $\Delta n = \pm 2$ and $\Delta n = \pm 4$ are shown in Figures
2f-k. The dominant contributions are the electron-like intraband transitions
$n\rightarrow n+2$ to $\sigma_{-}(\omega)$ and $n\rightarrow n+4$ to
$\sigma_{+}(\omega)$. Although one may expect these overtones to be much
weaker than the main cyclotron resonance, they have, in fact, a comparable
intensity and form outstanding satellite structures in the total conductivity
(Figures 2b,c). Qualitatively, one can relate this very peculiar observation
with the comparable values of $\gamma_3$ and other interlayer hopping
parameters, for example $\gamma_1\approx 0.4$~eV. Interestingly, the optical
intensities of these two harmonics are about the same, contrary to a naive
expectation of having a stronger transition for a smaller $\Delta n$. This
can be attributed to the fact that trigonal warping most strongly mixes LLs
separated by 3, thus making the harmonics for $\Delta n = 3\pm 1$ the
strongest and roughly equal in intensity. Our calculations show indeed that
the transitions of higher orders (5, 7, etc.) are significantly weaker than
the first two harmonics.

Figure 3 presents the evolution of the calculated optical conductivity
spectra as a function of the magnetic field. It is evident that the
intensities of the transitions $n\rightarrow n+2$ and $n\rightarrow n+4$ are
of the same order of magnitude as the one of the transitions $n\rightarrow
n+1$ up to at least 7~T and all of them originate from the Drude peak at zero
field. In fact, the same calculations (not shown) predict that the high-order
harmonics should remain optically visible up to extremely strong fields
$\sim$ 100~T. Importantly, such a strong effect of higher harmonics must also
exist in multilayer graphene with a similar (Bernal-type) stacking, where
trigonal warping is present~\cite{AbergelPRB07}.

To summarize, we demonstrated that a classical band model, if rigorously
applied, describes accurately the cyclotron spectra in graphite in a broad
range of magnetic fields. This should serve as a solid basis for the optical
investigation of more subtle phenomena such as the coupled electron-hole
plasma, electron-phonon interactions and the spin-orbit coupling, the effects
essential for the emerging fields of graphite(graphene)-based plasmonics and
spintronics. We anticipate that measuring the spectroscopic Kerr angle that
allows distinguishing the high-frequency dynamics of electrons and holes
without using the circular polarized radiation, will be highly useful in the
future studies of the topological insulators, intercalated graphites,
multilayer graphene and other materials of fundamental and practical
importance. Finally, our work suggests that high-order cyclotron resonances
should also be strong in the optical spectra of multilayer graphene up to
high fields and therefore should play a major role in the related
applications.

This research was supported by the Swiss National Science Foundation by the
grants 200021-120347, 200020-135085 and IZ73Z0-128026 (SCOPES program),
through the National Center of Competence in Research ``Materials with Novel
Electronic Properties-MaNEP". The authors acknowledge useful discussions with
L.A. Falkovsky and thank I. Crassee and A. Akrap for critically reading the
manuscript.


\newpage

\begin{center}{\bf \large Decrypting the cyclotron effect in graphite using Kerr rotation spectroscopy: Supplemental Material}
\end{center}
\begin{center}Julien Levallois, Micha\"{e}l Tran and Alexey B. Kuzmenko
\end{center}
\begin{center}
{\it D\'{e}partement de Physique de la Mati\`{e}re Condens\'{e}e, Universit\'{e} de Gen\`{e}ve, CH-1211 Gen\`{e}ve 4, Switzerland}
\end{center}

\setcounter{equation}{0}
\setcounter{figure}{0}

\section{Experiment}

Magneto-optical measurements were done on a large piece (about
7$\times 7$ mm$^2$) of highly ordered pyrolytic graphite (HOPG)
of the ZYA grade, the same as used in
Ref.~\onlinecite{KuzmenkoPRL08}. The misorientation of the
z-axis is smaller than 0.4$^o$ as the X-ray rocking curve
showed. The sample was mounted in a split-coil superconducting
magnet attached to a Fourier-transform spectrometer. A mercury
infrared light source, a Ge-coated mylar beamsplitter and a
liquid helium cooled bolometer detector were used. The magnetic
field was applied along the z-axis perpendicular to the sample
and almost parallel to the propagation of light. The average
angle of incidence is $8^{\circ}$.

The Kerr rotation angle was determined by placing a rotating
grid-wire gold polarizer before the sample and a fixed one just
after the sample (before the detector). Without the magnetic
field, the signal is at minimum when the two polarizers are
crossed. In our configuration the minimum value was less than
1\% of the maximum value (when the polarizers are parallel),
which allows an accurate polarimetric measurement.

In magnetic field, the minimum shifts by the amount of the Kerr
rotation. At every field value, we collected the reflected
spectra for a number of polarizer angles around the minimum.
The minimum position $\phi_{min}(\omega)$ was then determined
as a continuous spectrum. The Kerr rotation is experimentally
determined as
\begin{equation}
\theta_K(\omega,B)=\phi_{min}(\omega,B) - \phi_{min}(\omega,0),
\end{equation}
\noindent while the direction of the polarizer rotation was
chosen to match the physical definition given by Eq.~(1) of the
main text. We verified that
$\theta_K(\omega,-B)=-\theta_K(\omega,B)$ within our
experimental accuracy. Therefore, in order to reduce the
measurement noise, we eventually determined the Kerr angle
using the relation:
\begin{equation}
\theta_K(\omega,B)=[\phi_{min}(\omega,B) -
\phi_{min}(\omega,-B)]/2.
\end{equation}
\noindent The experimental precision is better than 1~mrad.

The reflectivity of the sample $R(\omega,B)$ was determined using a
triple-reference method. First, a ratio of the intensities of a
sample ($I_s$) and a gold mirror ($I_r$) was taken at a given
field B:
\begin{equation}
\rho_(\omega,B)=I_{s}(\omega,B)/I_{r}(\omega,B).
\end{equation}
\noindent The switching between the sample and the reference
was done with the aid of a computer controlled translation
stage. The same ratio $\rho(\omega,0)$ was taken at zero field.
The absolute reflectivity was determined using the formula:
\begin{equation}
R(\omega,B)=R(\omega,0)\rho(\omega,B)/\rho(\omega,0),
\end{equation}
\noindent where $R(\omega,0)$ is the zero field reflectivity,
measured on the same sample in a different cryostat allowing an
\emph{in-situ} gold evaporation without displacing the sample~\cite{KuzmenkoPRL08s}. Such a procedure eliminates the sensitivity of the detector to magnetic field, temporal drifts
and other systematic uncertainties.

\section{Theory}

In deriving the Landau levels in graphite, we essentially
follow the method of Nakao~\cite{NakaoJPSJ76s,NakamuraJPSJ76s}.
The main difference is that instead of adopting the original
SWMcC Hamiltonian, we use the tight-binding Hamiltonian with
equivalent parameters $\gamma_0,
\gamma_1,\gamma_2,\gamma_3,\gamma_4, \gamma_5$ and $\Delta$, in
order to make a direct connection to the recent literature on
single- and multilayer graphene~\cite{CastroNetoRMP09s}. The
relation between the two Hamiltonians and their parameter sets
is given, for example, in Ref.~\onlinecite{PartoensPRB06s}.

The crystal structure of Bernal stacked graphite is shown in
Fig.~1, where also the SWMcC hopping parameters $\gamma_0, ...,
\gamma_5$ are specified. Due to a strong electronic anisotropy,
the band dispersion close to the K-H line is customarily
treated by the tight-binding model in the $z$-direction, and by
linear expansion in the in-plane momentum ($k_x,k_y$). For each
value of $k_z$ one can thus define:
\begin{eqnarray}
\tilde{\gamma}_{1,3,4} &=& 2\gamma_{1,3,4}\cos(k_{z}d/2)\\
\tilde{\gamma}_{2,5} &=& 2\gamma_{2,5}\cos(k_{z}d)
\end{eqnarray}
\noindent and write down the $4\times4$ in-plane Hamiltonian in
the \{B1, A1, A2, B2\} basis:
\begin{equation}
\hat{H}=\left(
  \begin{array}{cccc}
    \tilde{\gamma}_{2}  &  \hbar v k_{-}                     & \hbar\tilde{v}_{4} k_{-}          & \hbar\tilde{v}_{3} k_{+} \\
    \hbar v k_{+}             & \tilde{\gamma}_{5} + \Delta  & \tilde{\gamma}_{1}           & \hbar\tilde{v}_{4} k_{-} \\
    \hbar\tilde{v}_{4} k_{+} & \tilde{\gamma}_{1}           & \tilde{\gamma}_{5}  + \Delta &  \hbar v k_{-}             \\
    \hbar\tilde{v}_{3} k_{-} & \hbar\tilde{v}_{4} k_{+}          & \hbar v k_{+}                     & \tilde{\gamma}_{2}
  \end{array}
\right)
\end{equation}

\noindent where $k_{\pm} = k_{x} \pm i k_{y}$, $v =
(3/2)(\gamma_0 a/\hbar)\approx 10^6$ m/s is the Fermi velocity
and where we also introduced $\tilde{v}_{3,4} = v
\tilde{\gamma}_{3,4}/\gamma_{0}$. The last SWMcC parameter
$\Delta$ stands for the on-site energy difference between the
non-equivalent atomic positions A and B. Here $a$ = 1.43~\AA\
is the carbon-carbon distance and $d$ = 6.7\~\AA\ is twice the
interlayer separation (Fig.~1).

\begin{figure}
\includegraphics[width=5cm]{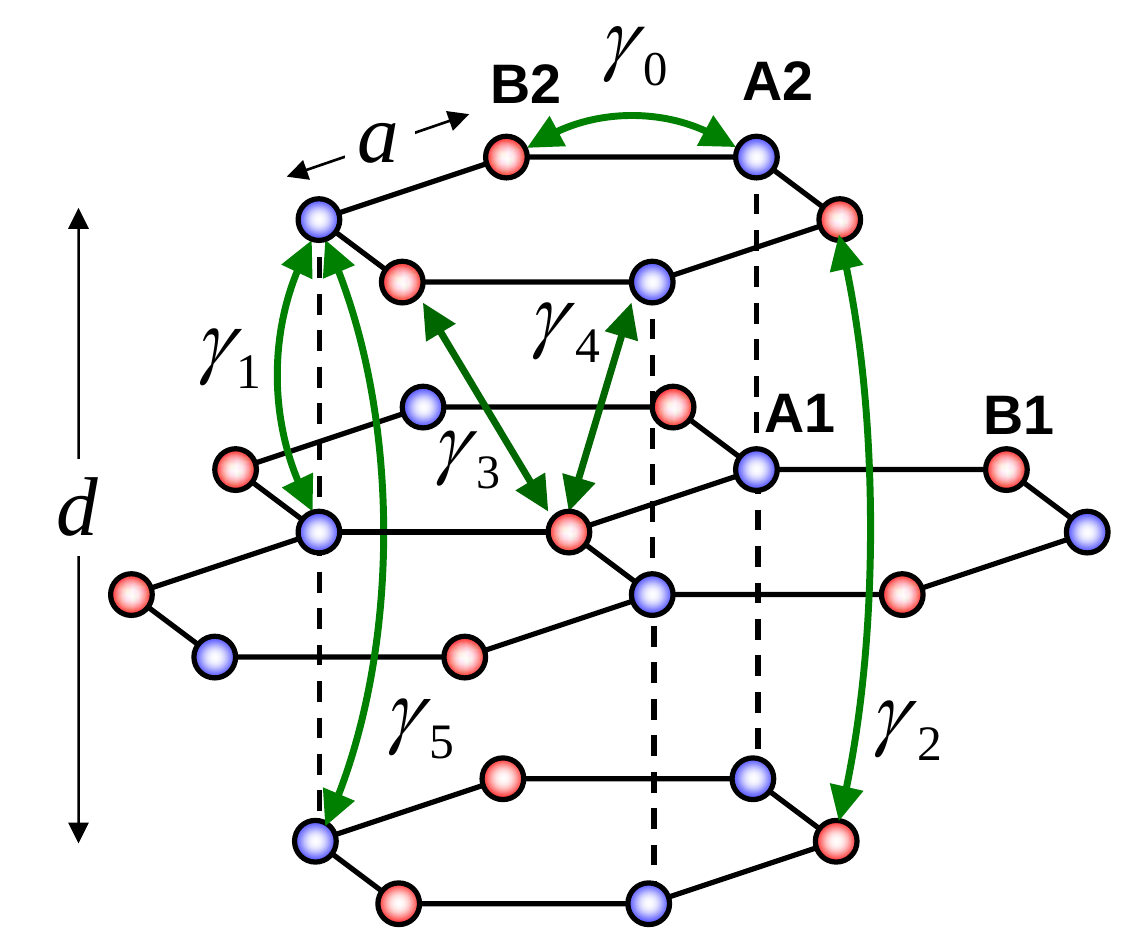}\\
\caption{(color online) Crystal structure of Bernal graphite with the
indication of the hopping parameters of the SWMcC model.}
\label{Fig2}
\end{figure}
The velocity operators for the right and the left polarizations
can be obtained straightforwardly:
\begin{equation}
\hat{v}_{+}=\frac{1}{\hbar}\frac{\partial{\hat{H}}}{\partial
k_{+}} = \left(
  \begin{array}{cccc}
    0             & 0             & 0  & \tilde{v}_{3} \\
    v             & 0             & 0  & 0             \\
    \tilde{v}_{4} & 0             & 0  & 0             \\
    0             & \tilde{v}_{4} & v  & 0
  \end{array}
\right),
\end{equation}

\begin{equation}
\hat{v}_{-}=\frac{1}{\hbar}\frac{\partial{\hat{H}}}{\partial
k_{-}} = \hat{v}_{+}^{*}.
\end{equation}

\noindent In a perpendicular magnetic field $B$, we make the
substitution~\cite{LuttingerPR55s,InoueJPSJ62s,AbergelPRB07s,FalkovskyPRB11s}:
\begin{eqnarray}
k_{-}&\rightarrow&a^{+}\sqrt{2eB/\hbar}  \\
k_{+}&\rightarrow&a\sqrt{2eB/\hbar},
\end{eqnarray}

\noindent where $a^{+}$ and $a$ are the ladder operators,
acting on harmonic oscillator-like wavefunctions $\psi_{n}$:
\begin{eqnarray}
a\psi_{n}     &=& \sqrt{n}\psi_{n-1}        \\
a^{+}\psi_{n} &=& \sqrt{n+1}\psi_{n+1}.
\end{eqnarray}

\noindent The Hamiltonian in the operator form reads:
\begin{equation}
\hat{H}=\left(
  \begin{array}{cccc}
    \tilde{\gamma}_{2} & \epsilon a^{+}              & \epsilon_{4} a^{+}          & \epsilon_{3} a     \\
    \epsilon a         & \tilde{\gamma}_{5} + \Delta & \tilde{\gamma}_{1}          & \epsilon_{4} a^{+} \\
    \epsilon_{4} a     & \tilde{\gamma}_{1}          & \tilde{\gamma}_{5} + \Delta & \epsilon a^{+}     \\
    \epsilon_{3} a^{+} & \epsilon_{4} a              &  \epsilon a                 & \tilde{\gamma}_{2}
  \end{array}
\right),
\end{equation}

\noindent where we introduced the cyclotron energies:
\begin{eqnarray}
\epsilon &=& v        \sqrt{2e\hbar B}\nonumber\\
\epsilon_{3,4} &=& \epsilon (v_{3,4}/v)\nonumber.
\end{eqnarray}

The states $\psi_{0}, \psi_{1},...$ form a complete orthogonal
basis on one sublattice. As in graphite there are four atoms
per unit cell, we add the sublattice index $m = 1, 2, 3, 4$ and
define the following set of basis states:
\begin{eqnarray}
\Psi_{n}^{m} = \left(
  \begin{array}{c}
    \delta_{1m} \\
    \delta_{2m} \\
    \delta_{3m} \\
    \delta_{4m}
  \end{array}
\right) \psi_{n}.
\end{eqnarray}

\noindent It is useful to group them in a particular
way~\cite{NakaoJPSJ76s}, namely: $\Gamma_{0} = \{\Psi_{0}^{1}\}$,
$\Gamma_{1} = \{\Psi_{1}^{1},\Psi_{0}^{2},\Psi_{0}^{3}\}$ and
$\Gamma_{n} =
\{\Psi_{n}^{1},\Psi_{n-1}^{2},\Psi_{n-1}^{3},\Psi_{n-2}^{4}\}$
for $n>1$ and arrange in the order $\Gamma_{0},
\Gamma_{1},\Gamma_{2},..$. Then the Hamiltonian can be written
as a matrix:
\begin{equation}
\hat{H}=\left(
  \begin{array}{cccccc}
    H_{0}     & 0         & 0      & G_{0}  & 0      & \cdots \\
    0         & H_{1}     & 0      & 0      & G_{1}  & \cdots \\
    0         & 0         & H_{3}  & 0      & 0      & \cdots \\
    G_{0}^{*} & 0         & 0      & H_{4}  & 0      & \cdots \\
    0         & G_{1}^{*} & 0      & 0      & H_{5}      & \cdots \\
    \vdots    & \vdots    & \vdots & \vdots & \vdots & \ddots
  \end{array}
\right),\label{EqH}
\end{equation}
\noindent where:
\begin{eqnarray}
&&H_{0}=\left(
  \begin{array}{c}
    \tilde{\gamma}_{2}
  \end{array}
\right), H_{1}=\left(
  \begin{array}{ccc}
    \tilde{\gamma}_{2} & \epsilon           & \epsilon_{4}          \\
    \epsilon           & \tilde{\gamma}_{5} & \tilde{\gamma}_{1}    \\
    \epsilon_{4}       & \tilde{\gamma}_{1} & \tilde{\gamma}_{5}
  \end{array}
\right),\nonumber\\
&&H_{n=2,3...}=\left(
  \begin{array}{cccc}
    \tilde{\gamma}_{2}     & \epsilon\sqrt{n}            & \epsilon_{4}\sqrt{n}        & 0                      \\
    \epsilon\sqrt{n}       & \tilde{\gamma}_{5} + \Delta & \tilde{\gamma}_{1}          & \epsilon_{4}\sqrt{n-1} \\
    \epsilon_{4}\sqrt{n}   & \tilde{\gamma}_{1}          & \tilde{\gamma}_{5} + \Delta & \epsilon \sqrt{n-1}    \\
    0                      & \epsilon_{4}\sqrt{n-1}      &  \epsilon \sqrt{n-1}        & \tilde{\gamma}_{2}
  \end{array}
\right),\nonumber\\
&&G_{0}=\left(
  \begin{array}{cccc}
    0 & 0 & 0 & \epsilon_{3}
  \end{array}
\right), G_{1}=\left(
  \begin{array}{cccc}
    0 & 0 & 0 & \epsilon_{3}\sqrt{2} \\
    0 & 0 & 0 & 0                    \\
    0 & 0 & 0 & 0
  \end{array}
\right),\nonumber\\
&&G_{n=2,3...}=\left(
  \begin{array}{cccc}
    0 & 0 & 0 & \epsilon_{3}\sqrt{n+1} \\
    0 & 0 & 0 & 0                      \\
    0 & 0 & 0 & 0                      \\
    0 & 0 & 0 & 0
  \end{array}
\right).\nonumber
\end{eqnarray}

The Landau levels are calculated by the diagonalization of the
matrix (\ref{EqH}). In the case $\gamma_3 = 0$ (no trigonal
warping) all $G_{n}$ vanish and the Hamiltonian can be
factorized into individual blocks $H_{n}$ each of those can be
solved separately to provide the sublevels. If $\gamma_{3} \neq
0$ then one has to diagonalize the infinite Hamiltonian, which
can be further factorized~\cite{NakaoJPSJ76s} into three
independent, although still infinite, matrices by separating
subsets $A=\{\Gamma_{0}, \Gamma_{3},\Gamma_{6},..\}$,
$B=\{\Gamma_{1}, \Gamma_{4},\Gamma_{7},..\}$ and
$C=\{\Gamma_{2}, \Gamma_{5},\Gamma_{8},..\}$. Numerically, the
problem can be tackled by truncating an infinite matrix to
obtain a finite square matrix $M\times M$ and making $M$ large
enough so that increasing it further changes negligibly the
energies and the eigenfunctions of the low-index Landau levels
of interest. The bands resulting from subsets A, B and C are
marked in Fig.2a of the main text by black, red and blue colors
respectively.

The velocity operator in the same representation is
\begin{equation}
\hat{v}_{+}=\hat{v}_{-}^{*}=\left(
  \begin{array}{cccccc}
    0      & 0      & U_{0}  & 0      & 0      & \cdots \\
    V_{1}  & 0      & 0      & U_{1}  & 0      & \cdots \\
    0      & V_{2}  & 0      & 0      & U_{2}  & \cdots \\
    0      & 0      & V_{3}  & 0      & 0      & \cdots \\
    \vdots & \vdots & \vdots & \vdots & \vdots & \ddots
  \end{array}
\right),
\end{equation}
\noindent where
\begin{eqnarray}
&&V_{1}=\left(
  \begin{array}{c}
    0             \\
    v             \\
    \tilde{v}_{4} \\
  \end{array}
\right), V_{2}=\left(
  \begin{array}{ccc}
    0             & 0             & 0 \\
    v             & 0             & 0 \\
    \tilde{v}_{4} & 0             & 0 \\
    0             & \tilde{v}_{4} & v
  \end{array}
\right),\nonumber\\
&&V_{3,4...}=\left(
  \begin{array}{cccc}
    0             & 0             & 0  & 0 \\
    v             & 0             & 0  & 0 \\
    \tilde{v}_{4} & 0             & 0  & 0 \\
    0             & \tilde{v}_{4} & v  & 0
  \end{array}
\right),\nonumber\\
&&U_{0}=\left(
  \begin{array}{cccc}
    0             & 0             & 0  & \tilde{v}_{3}
  \end{array}
\right), U_{1}=\left(
  \begin{array}{cccc}
    0             & 0             & 0  & \tilde{v}_{3} \\
    0             & 0             & 0  & 0             \\
    0             & 0             & 0  & 0
  \end{array}
\right),\nonumber\\
&&U_{2,3...}=\left(
  \begin{array}{cccc}
    0             & 0             & 0  & \tilde{v}_{3} \\
    0             & 0             & 0  & 0             \\
    0             & 0             & 0  & 0             \\
    0             & 0             & 0  & 0
  \end{array}
\right).\nonumber
\end{eqnarray}

The circular optical conductivity is calculated using the Kubo
formula (the spin and the valley degeneracies are taken into
account):
\begin{eqnarray}
\sigma_{\pm}(\omega)=\frac{8e^3
B}{\pi^{2}}\int_{0}^{\frac{\pi}{d}}d k_{z}\sum_{i,j\neq
i}\left|\left\langle
i\left|\hat{v}_{\pm}\right|j\right\rangle\right|^2\nonumber\\
\times\frac{f(\epsilon_{i}) - f(\epsilon_{j})}{\epsilon_{j} -
\epsilon_{i}} \frac{i}{\hbar\omega -
\epsilon_{j}+\epsilon_{i}+i\Gamma}.\label{EqSigma}
\end{eqnarray}
\noindent The indices $i$ and $j$ run over all Landau levels at
a given $k_z$. Here
$f(\epsilon)=(1+\exp{(\epsilon-\mu)/T})^{-1}$ is the Fermi
distribution function. The chemical potential $\mu$ depends on
the magnetic field and can be adjusted using the electrical
neutrality condition ~\cite{KuzmenkoPRB09s}:
\begin{eqnarray}
\int_{0}^{\frac{\pi}{d}}d
k_{z}\sum_{i}\left[f(\epsilon_{i})-1/2\right] = 0.
\end{eqnarray}

The optical longitudinal and Hall conductivities are derived
from the circular one:
\begin{eqnarray}
\sigma_{xx}(\omega)&=&[\sigma_{+}(\omega)+\sigma_{-}(\omega)]/2\nonumber\\
\sigma_{xy}(\omega)&=&[\sigma_{+}(\omega)-\sigma_{-}(\omega)]/2i\nonumber.
\end{eqnarray}

In the absence of trigonal warping, only the matrix elements
between the levels which quantum numbers $n_i$ and $n_j$ differ
by $\pm$ 1 are non-zero. If $\gamma_3 \neq 0$ then in principle
all transitions are allowed except those where $n$ changes by
$0, \pm3, \pm6...$, i.e. when both LLs originate from the same
subset (A, B or C). As our calculation shows (see the main
text) the transitions with $\Delta n = \pm 2$ and $\pm 4$ have
the strongest optical intensity, while the others (for example
$\Delta n = \pm 5$ and $\pm 7$) are much weaker.

In reality one has to limit the summation to a finite number of
LL transitions. Since our goal is to calculate optical
properties in a limited spectral range $0<\omega<\omega_{max}$,
it is natural to keep only the resonances below a certain
threshold value $\Omega > \omega_{max}$. Obviously, the total
number of counted transitions then depends on the magnetic
field.

The reflectivity and the Kerr angle are given by the formulas
\begin{eqnarray}
R &=& \frac{|r_{+}|^2 + |r_{-}|^2}{2}\label{EqR}\\
\theta_{K} &=& \frac{\arg(r_{-}) - \arg(r_{+})}{2}\label{EqKerr},
\end{eqnarray}

\noindent where $r_{\pm}$ are the complex reflectivities for
the circularly polarized light. At a normal incidence, they are
given by the Fresnel equation:
\begin{equation}
r_{\pm} = \frac{1 - \sqrt{\epsilon_{\pm}}}{1 +
\sqrt{\epsilon_{\pm}}},\label{EqRPM}
\end{equation}

\noindent where
\begin{equation}
\epsilon_{\pm}(\omega) = \epsilon_{\infty} +
\frac{4\pi\sigma_{\pm}(\omega)i}{\omega}.\label{EqEpsilonPM}
\end{equation}
is the in-plane dielectric function. The purpose of the
parameter $\epsilon_{\infty}$ is to account for the static
dielectric constant associated with all higher-energy optical
transitions, not covered by Eq.~(\ref{EqSigma}). It should not
be considered as a free parameter, since it is determined by
the choice of $\Omega$ and can be fixed using the
experimentally obtained dielectric function at zero field~\cite{KuzmenkoPRL08s}.

The integration over $k_z$ in practice is done by a summation
over a finite grid. One has to choose it in such a way that
making it denser does not change the result anymore. The
optimal density depends on scattering ($\Gamma$), temperature
($T$) and field ($B$).

\section{Fitting procedure and results}

We fitted the Kerr spectra at 1, 3, 5 and 7 T simultaneously
with the Equations (\ref{EqKerr}), (\ref{EqRPM}),
(\ref{EqEpsilonPM}) and (\ref{EqSigma}), by adjusting the SWMcC
parameters except $\gamma_0$ that we fixed to a value
corresponding to $v = 1.02\times 10^6$ m/s
(Ref.~\onlinecite{OrlitaPRL08s}). All parameters are assumed to
be field-independent. The least-squares criterion was used and
the Levenberg-Marquardt fitting routine was employed~\cite{NumericalRecipess}, with an analytical calculation of the
derivatives of the output with respect to parameters. The
RefFIT software~\cite{Reffits} was used with a dedicated model
for graphite. Only a few iterations were needed to achieve a
convergence. The best values are presented in Table~\ref{TableFit}, where also the results of the fitting of recent magnetotransport measurements~\cite{SchneiderPRL09s} are shown,
as well as the SWMcC parameters obtained by fitting the
calculated bands using the first-principle GW method~\cite{GruneisPRB08s}. One can see that our parameter values agree well with the previous works. The relatively large error bars
for $\gamma_4$, $\gamma_5$ and $\Delta$ are due to the fact
that their effects on optical spectra are correlated. By fixing
one of them to other experiments, or using magneto-optical data
at higher frequencies and magnetic fields~\cite{LiPRB06s,TungCM11s} we can likely reduce the error bars.
However, the main goal of the present work is to demonstrate
that the Kerr spectra are fully consistent with the SWMcC
model.

\begin{figure}
\includegraphics[width=5cm]{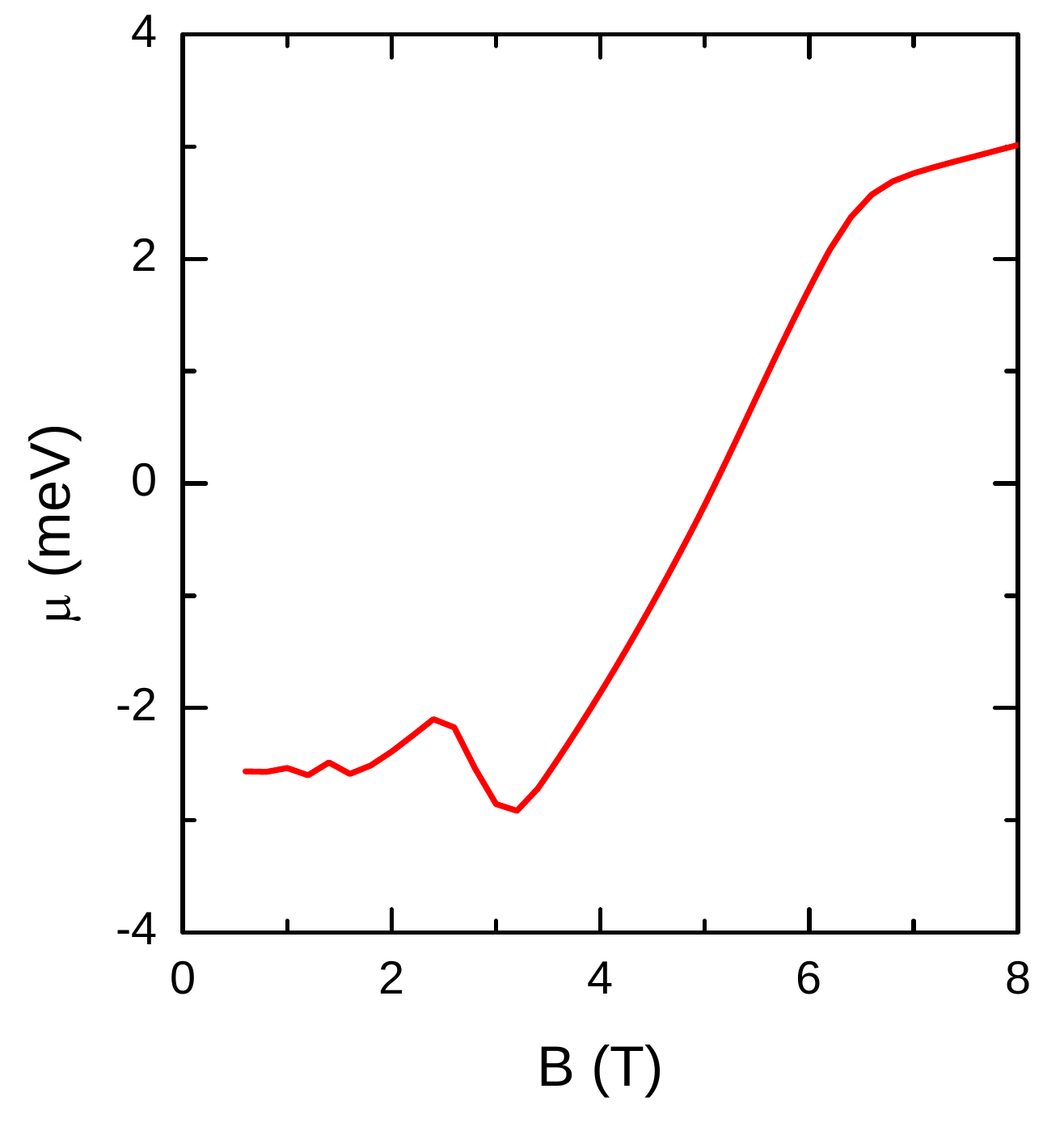}\\
\caption{(color online) Magnetic field dependence of the chemical potential.}
\label{FigChemPot}
\end{figure}

Fig.~\ref{FigChemPot} shows the dependence of the chemical
potential on magnetic field. The non-monotonic dependence is
due to the crossing of the Fermi level by the LLs. Almost
identical curve is obtained in Ref.~\onlinecite{SchneiderPRL09s}.

\begin{table}
\caption{SWMcC parameters (in eV) obtained in the present work
(1st column), in Ref.~\onlinecite{SchneiderPRL09s} (2nd column) and
Ref.~\onlinecite{GruneisPRB08s} (3rd column).}

\begin{tabular}{llll}
  \hline
  \hline
  \ \ \ \ \ & This work & Schneider et al. ~\cite{SchneiderPRL09s} & Gruneis et al.~\cite{GruneisPRB08s}\\
  \hline
  $\gamma_{0}$ &  3.16   (fixed)      &  3.37   $\pm$ 0.02    &  3.05   \\
  $\gamma_{1}$ &  0.38  $\pm$ 0.01   & 0.363   $\pm$ 0.05    &  0.403  \\
  $\gamma_{2}$ & -0.0089  $\pm$ 0.0003 & -0.0121 $\pm$ 0.0005  & -0.0125 \\
  $\gamma_{3}$ &  0.297   $\pm$ 0.005  &  0.31   $\pm$ 0.05    &  0.274  \\
  $\gamma_{4}$ & -0.15   $\pm$ 0.1    & -0.07   $\pm$ 0.01    & -0.143  \\
  $\gamma_{5}$ &  0.005  $\pm$ 0.03   &  0.025  $\pm$ 0.005   &  0.015  \\
  $\Delta$     &  0.027   $\pm$ 0.05   &  0.024  $\pm$ 0.002   &  0.05   \\
  $\Gamma$     &  0.0028 $\pm$ 0.0002   &                     &         \\
  \hline
  \hline
\end{tabular}
\label{TableFit}

\end{table}

In the calculations, both $\Omega$ and $\epsilon_{\infty}$ were
the same at all fields (0.2~eV and 15 respectively). The size
of the truncated matrix $M$ and the number of $k_z$ grid points
$N_{z}$ was optimized for every field in order to speed up the
calculations while keeping a sufficient accuracy. The actual
values used are given in Table II.

\begin{table}
\caption{The used computational settings for every field value
as explained in the text.}
\begin{tabular}{cccccccccc}

\hline
\hline
$B$ (T) & 0.5 & 1 & 1.5 & 2 & 3 & 4 & 5 & 6 & 7\\
\hline
$M$ & 180 & 90 & 60 & 45 & 30 & 24 & 18 & 15 & 15\\

$N_{z}$ & 50 & 100 & 150 & 200 & 300 & 400 & 500 & 600 & 700\\
  \hline
  \hline
\end{tabular}
\label{TableFit}

\end{table}

\begin{figure}
\includegraphics[width=8cm]{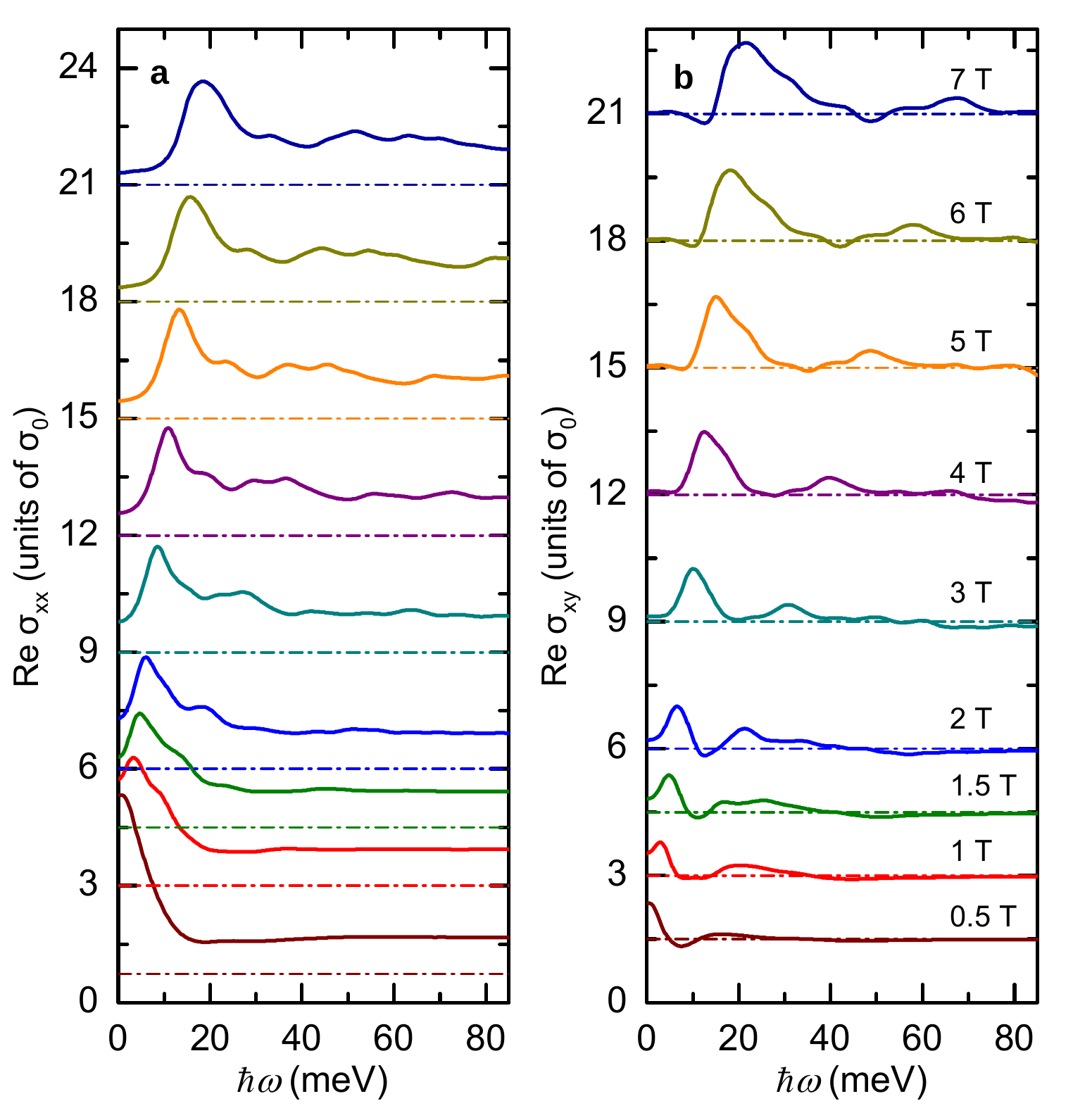}\\
\caption{(color online) Real parts of $\sigma_{xx}(\omega)$ (a) and $\sigma_{xy}(\omega)$ (b) that correspond to $\sigma_{\pm}(\omega)$ of Fig.3 of the main text.}
\label{FigSxxSxy}\end{figure}

For completeness, in Fig.~\ref{FigSxxSxy} we present the
longitudinal and the Hall conductivities obtained from the same
calculation (Fig.~3 of the main text).


\begin{thebibliography}{99}

\bibitem{GaltPR56}
J. K. Galt, W. A. Yager, and H. W. Dail, Jr., Phys. Rev. \textbf{103}, 1586
(1956).

\bibitem{SuematsuJPSJ72}
H. Suematsu and S. Tanuma, J. Phys. Soc. Japan. \textbf{33}, 1619 (1972).

\bibitem{ToyPRB77}
W. W. Toy, M. S. Dresselhaus and G. Dresselhaus, Phys. Rev. B \textbf{15}, 4077 (1977).

\bibitem{DoezemaPRB79} R.E. Doezema, W.R. Datars, H. Schaber, and A. Van Schyndel, Phys. Rev. B
    \textbf{19}, 4224 (1979).

\bibitem{NakamuraJPSJ84} K. Nakamura \emph{et al.},
    J.Phys. Soc. Japan. \textbf{53}, 1164 (1984).

\bibitem{HofmannRSI06} T. Hofmann \emph{et al.}, Rev. Sci. Instr.
    \textbf{77}, 063902 (2006).

\bibitem{LiPRB06} Z. Q. Li \emph{et al.}, Phys. Rev. B \textbf{74}, 195404
    (2006).

\bibitem{OrlitaPRL08} M. Orlita \emph{et al.}, Phys. Rev. Lett. \textbf{100},
    136403
    (2008).

\bibitem{ChuangPRB09}
K.-C. Chuang, A. M. R. Baker, and R. J. Nicholas, Phys. Rev. B \textbf{80}, 161410(R) (2009).

\bibitem{UbrigPRB11} N. Ubrig \emph{et al.}, Phys. Rev. B \textbf{83},
    073401 (2011).

\bibitem{TungCM11} L.-C. Tung, \emph{et al.}, arXiv:1106.5948 (2011).

\bibitem{GusyninPRL07} V. P. Gusynin, S. G. Sharapov, and J. P. Carbotte,
    Phys. Rev. Lett. \textbf{98}, 157402 (2007).

\bibitem{LiNP08} Z. Q. Li \emph{et al.}, Nature Physics \textbf{4}, 532
    (2008).

\bibitem{CrasseePRB11} I. Crassee \emph{et al.},
    Phys. Rev. B \textbf{84}, 035103 (2011).

\bibitem{CrasseeNP11} I. Crassee \emph{et al.}, Nature Physics \textbf{7}, 48
    (2011).

\bibitem{KaplanPRL96}
S. G. Kaplan et al., Phys. Rev. Lett. \textbf{76}, 696 (1996).

\bibitem{McClurePR57} J.W. McClure, Phys. Rev. \textbf{108}, 612 (1957).

\bibitem{SlonczewskiPR58} J.C. Slonczewski and P.R. Weiss, Phys. Rev.
    \textbf{109}, 272 (1958).

\bibitem{AbergelPRB07}
D. S. L. Abergel and V. I. Fal'ko, Phys. Rev. B \textbf{75}, 155430 (2007).

\bibitem{FalkovskyPRB11}
L.A. Falkovsky, Phys. Rev. B \textbf{83}, 081107 (2011).

\bibitem{SchneiderPRL09} J. M. Schneider, M. Orlita, M. Potemski, and D.K. Maude, Phys. Rev. Lett.
    \textbf{102}, 166403 (2009).

\bibitem{ZhuNP09} Z. Zhu \emph{et al.}, Nature Physics \textbf{6}, 26 (2009).

\bibitem{GruneisPRB08} A. Gruneis \emph{et al.}, Phys. Rev. B \textbf{78},
    205425 (2008).

\bibitem{NakaoJPSJ76}
K. Nakao, J. Phys. Soc. Japan. \textbf{40}, 761 (1976).

\bibitem{KoshinoPRB08}
M. Koshino and T. Ando, Phys. Rev. B \textbf{77}, 115313 (2008).

\bibitem{NozieresPR58}
P. Nozieres, Phys. Rev. \textbf{109}, 1510 (1958).

\bibitem{InoueJPSJ62}
M. Inoue, J. Phys. Soc. Japan. \textbf{17}, 808 (1962).

\bibitem{MuchaPRB11}
M. Mucha-Kruczynski, I.L. Aleiner, V.I. Fal'ko, Phys. Rev. B \textbf{84}, 041404 (2011).

\bibitem{NovoselovScience11} A. S. Mayorov \emph{et al.}, Science
    \textbf{333}, 860 (2011).

\bibitem{KuzmenkoPRL08} A. B. Kuzmenko, E. van Heumen, F. Carbone, and D. van der Marel, Phys. Rev. Lett.
    \textbf{100},
    117401 (2008).

\end{thebibliography}

\begin{thebibliography}{99}

\bibitem{KuzmenkoPRL08s}
A. B. Kuzmenko \emph{et al.}, Phys. Rev. Lett. \textbf{100}, 117401 (2008).

\bibitem{NakaoJPSJ76s}
K. Nakao, J. Phys. Soc. Japan. \textbf{40},
761 (1976).

\bibitem{NakamuraJPSJ76s}
K. Nakamura \emph{et al.}, J. Phys. Soc.
Japan. \textbf{53}, 1164 (1983).

\bibitem{CastroNetoRMP09s}
A.H. Castro Neto et al., Rev. Mod. Phys. \textbf{81}, 109
(2009).

\bibitem{PartoensPRB06s}
B. Partoens and F.M. Peeters, Phys. Rev. B {\bf 74}, 075404
(2006).

\bibitem{LuttingerPR55s}
J.M. Luttinger and W. Kohn, Phys. Rev. {\bf 97}, 869 (1955).

\bibitem{InoueJPSJ62s}
M. Inoue, J. Phys. Soc. Japan. \textbf{17}, 808 (1962).

\bibitem{AbergelPRB07s}
D. S. L. Abergel and V. I. Fal'ko, Phys. Rev. B \textbf{75},
155430 (2007).

\bibitem{FalkovskyPRB11s}
L.A. Falkovsky, Phys. Rev. B \textbf{83}, 081107 (2011).

\bibitem{KuzmenkoPRB09s}
A. B. Kuzmenko \emph{et al.}, Phys. Rev. B \textbf{80}, 165406 (2009).

\bibitem{OrlitaPRL08s}
M. Orlita \emph{et al.}, Phys. Rev. Lett. \textbf{100}, 136403 (2008).

\bibitem{NumericalRecipess}
W. H. Press, B. P. Flannery, S. A. Teukolsky, and W. T.
Vetterling, \emph{Numerical Recipes in C}, Cambridge Univ.
Press (1986).

\bibitem{Reffits}
A. B. Kuzmenko, {\it Guide to RefFIT: software to fit optical
spectra}, available online:
http://optics.unige.ch/alexey/reffit.html.

\bibitem{SchneiderPRL09s}
J. M. Schneider \emph{et al.}, Phys. Rev. Lett. \textbf{102}, 166403
(2009).

\bibitem{GruneisPRB08s}
A. Gruneis \emph{et al.}, Phys. Rev. B \textbf{78}, 205425 (2008).

\bibitem{LiPRB06s}
Z. Q. Li \emph{et al.}, Phys. Rev. B \textbf{74}, 195404 (2008).

\bibitem{TungCM11s}
L.-C. Tung \emph{et al.}, arXiv:1106.5948 (2011).

\end{thebibliography}
\end{document}